\documentclass[12pt]{article}
\usepackage{amsmath,amsfonts,amssymb,cite}
\begin{document}
\renewcommand{\thefootnote}{\fnsymbol{footnote}} 
\begin{center}
{\Large\bf QCD tests for Quasilocal Quark Model}\footnote{Talk given at
12th International Seminar on High Energy Physics QUARKS'2002
                        Novgorod, Russia, June 1-7, 2002} \\
\vspace{4mm}

{\large A. A. Andrianov}$^{\sharp\diamond}$, 
{\large V. A. Andrianov}$^{\sharp}$ and {\large S. S. Afonin}$^{\sharp}$ \\
$^{\sharp}$V.A.Fock Department of Theoretical Physics, St.Petersburg State
University, Russia\\
$^{\diamond}$ Istituto Nazionale di Fisica Nucleare, Sezione di Bologna, Italy\\

\bigskip

{\bf Abstract}
\end{center}

{\small We
perform the QCD testing of the Quasilocal
Quark Model (QQM) based on Operator Product
Expansion (OPE). The quark current
correlators calculated in framework of the model, 
are compared to their OPE in QCD at intermediate energies. 
The QQM provides a reasonable
resolution for mass spectrum of  parity doublers 
in scalar and vector meson channels.}\\

{\large\bf 1. Introduction}\\

Effective quark models are widely used to simulate main
features of nonperturbative QCD at low and intermediate energies
while having advantages to be more tractable in calculations. 
The quality of a simulation has to be  controlled
by a number of QCD tests. First of all, an effective
model should reproduce the symmetries of QCD. Second, the chiral and
conformal symmetries are broken in QCD eventually leading
to formation of quark and gluon condensates respectively.
The latter one should be also embedded into a model. Third, quark current
correlators calculated in framework of a model, 
are to be matched to the Operator Product
Expansion (OPE)~\cite{svz} at intermediate energies. 
One could mention also the heavy mass matching for $m_q\gg \Lambda_{QCD}$\,
and the reproduction of chiral and scale anomalies which however are not 
involved in our
discussion.

The requirements enumerated above are quite severe. For example,
the Chiral Perturbation Theory containing the pseudoscalar degrees
of freedom only does not pass all the QCD tests: matching to the
OPE gives wrong results for this theory. In the present report we
perform the QCD testing of the so called Quasilocal
Quark Model (QQM), which admits as linear realization~\cite{a1, a2, a3} as
non-linear one~\cite{ae1, ae2, ae3}. It fulfils the
matching to the OPE rather successfully. 

In this talk we 
deal with the linear realization of QQM which 
represents an extension of NJL-type
models~\cite{1,2,4,6,9,10} and 
allows to describe not only ground states of scalar ($S$),
pseudoscalar ($P$), vector ($V$), and axial-vector ($A$) mesons but also
their radial excitations known from Particle Data~\cite{pdg}.

The minimal structure of $SP$, $SU(2)$ QQM was discussed
in~\cite{a1, a3} and the $V\!A$, $SU(2)$ case was outlined in~\cite{we1}.
Here we consider the $SPV\!A$, $U(3)$ QQM~\cite{we2}.
In the Euclidean
space the relevant Lagrangian has the form:
\begin{eqnarray}
L&=&i\bar{q}\left(\hat{\not{\partial\phantom{d}}}\!\!\!+\hat{m}\right)q +
\frac{1}{4N_{f}N_{c}\Lambda^{2}}\sum_{k,l=1}^{2}\mbox{\rm Tr}\left\{
a_{kl}^a\sum_{j=1}^2\bar{q}f_{k}(\tau)\Gamma_j^aq\bar{q}f_{l}(\tau)\Gamma_j^aq
\right.\nonumber\\ 
&&\left.+
b_{kl}^a\sum_{j=3}^4\bar{q}f_{k}(\tau)\Gamma_j^aq\bar{q}f_{l}(\tau)\Gamma_j^aq
\right\}.
\end{eqnarray}
$a_{kl}^a$ and $b_{kl}^a$ represent here symmetric matrices of
real coupling constants in $SP$ and $V\!A$ case respectively.
Symbols $\Gamma_j^a$ mean:
\begin{equation}
\Gamma_1^a\equiv \lambda^a,\quad
\Gamma_2^a\equiv i\gamma_5\lambda^a,\quad
\Gamma_3^a\equiv i\gamma_{\mu}\lambda^a,\quad
\Gamma_4^a\equiv i\gamma_5\gamma_{\mu}\lambda^a;\qquad
a=0,...,8\,,
\end{equation}
where
\begin{equation}
\lambda^a=\frac{1}{\sqrt{2}}\lambda_{G-M}^a\,,\quad a=1,...,7\,,
\end{equation}
\begin{equation}
\lambda^0=\frac{\lambda_{G-M}^0+\lambda_{G-M}^8}{\sqrt{6}}\,,\quad
\lambda^8=\frac{-\lambda_{G-M}^0+\sqrt{2}\,\lambda_{G-M}^8}{\sqrt{6}}\,,
\end{equation}
with $\lambda_{G-M}^a$ being the standard set of Gell-Mann
matrices. The current quark mass matrix is
$\hat{m}=diag(m_u,m_d,m_s)$.
In the sequel we adopt the exact isospin symmetry $m_u=m_d$.
The symbol $\tilde{u}$ will stand everywhere for the $u$, $d$, $\bar{u}$,
$\bar{d}$ quarks. The symbol $\tilde{s}$ will denote $s$ or $\bar{s}$
quarks. We choose the polynomial
form-factors to be orthogonal on the unit interval:
\begin{equation}
\int_{0}^{1}f_{k}(\tau)f_{l}(\tau)d\tau=\delta_{kl}\,,
\end{equation}
\begin{equation}
f_{1}(\tau)=2-3\tau; \qquad f_{2}(\tau)=-\sqrt{3}\,\tau; \qquad
\tau\equiv-\frac{\partial^{2}}{\Lambda^{2}}\,.
\end{equation}
The parameter $\Lambda$ is a  four-momentum cutoff for virtual quark
momenta in quark loops. $N_c$ denotes a number of colors and $N_f=3$ is
the number of quark flavors.

Let us comment the approximations which will be used to
derive the meson characteristics: namely, the large $N_c$
and leading-log ($\ln\!\frac{\Lambda^2}{\mu^2}\gg1$) approximations.
The first one is equivalent~\cite{hooft1974,witten1979} to 
neglecting of meson loops. The second one fits well the quarks
confinement  as
quark-antiquark threshold contributions are suppressed in two-point
functions in the leading-log approximation. 
The accuracy of this approximation is controlled by the magnitudes
of higher dimensional
operators neglected in QQM, i.e. by contributions of heavy
mass resonances not included into QQM. All these approximations
are mutually consistent. 

We will work with bosonized action. Thus, one introduces auxiliary
$SPV\!A$-fields following the standard procedure:
\begin{equation}
L_{aux}=i\bar{q}\left(\hat{\not{\partial\phantom{d}}}\!\!\!+
\hat{m}+\sum_{k=1}^{2}\varphi_{k,j}^a\Gamma_j^a f_k(\tau)\right)q+
N_{c}N_f\Lambda^{2}\sum_{k,l=1}^{2}Tr\,\varphi_{k,j}^a
\left(c_{kl}^a\right)^{-1}\varphi_{l,j}^a\,,
\end{equation}
where $\varphi^a\equiv\sigma^a,\pi^a,\rho^a,A^a$ represents
auxiliary $S,P,V,A$ fields and $c_{kl}^a$ denotes $a_{kl}^a$ for
$SP$-case and $b_{kl}^a$ for $V\!A$-case.

For the cancellation of quadratic divergences of order~$\Lambda^2$ the
following parameterization of coupling constants is accepted in the
$SP$-case:
\begin{equation}
\label{par1}
8\pi^{2}\left(a_{kl}^a\right)^{-1}=
\delta_{kl}-\frac{\Delta_{kl}^a}{\Lambda^{2}};
\qquad \Delta_{kl}\ll\Lambda^{2}\,,
\end{equation}
where the physical mass parameters $\Delta_{kl}^a$ satisfy the
relations:
\begin{equation}
\label{par2}
\Delta_{kl}^m=\Delta_{kl}^0\,,\quad\Delta_{kl}^n=\frac12\left(\Delta_{kl}^0+
\Delta_{kl}^8\right)\,;\qquad m=1,2,3\,,\quad n=4,5,6,7\,.
\end{equation}
The same conditions are valid for $V\!A$-case with the replacement
\begin{equation}
\left(a_{kl}^a\right)^{-1}\rightarrow2\left(b_{kl}^a\right)^{-1},\qquad
\Delta_{kl}^a\rightarrow\frac43\bar\Delta_{kl}^a.
\end{equation}
The self-consistency of the mass spectrum turns out
to impose the following scale conditions:
\begin{equation}
\Delta_{kl}^i\sim \ln\!\frac{\Lambda^2}{M_0^2}\,;\quad
\bar\Delta_{kl}^i\sim\Lambda^2\,;\quad m_q^i\Lambda^2\sim1\,,
\end{equation}
where $M_0$ is a dynamic quark mass.

\bigskip

{\large\bf 2. Mass formulas from QQM with account of OPE}\\

The expressions for the mass spectrum are
displayed in~\cite{we2}. Here we present the mass relations, which are
independent of model parameters in the large-log approximation.
Some comments are in order.
The gluon anomaly is omitted in the
present report. Thus, the $\eta'$-meson is not considered. However,
we do not see any reason for appearance of $U_A(1)$ problem for
excited states. Below on the prime denotes excited states and the
symbol $(\eta)'$ means the first radial excitation of
$\eta$-meson.

First of all, the Gell-Mann-Okubo relation holds in the model and
the singlet state has no admixture of $s$-quark in $U(3)$ case:
\begin{equation}
\label{gmo}
m_{\alpha,\tilde{u}\tilde{u}}^2+m_{\alpha,\tilde{s}\tilde{s}}^2=
2m_{\alpha,\tilde{s}\tilde{u}}^2\,;\qquad
m_{\alpha,\tilde{u}\tilde{u}}=m_{\alpha,singlet}\,.
\end{equation}
Here $\alpha\equiv S,V,A;S',V',A',P'$. For
$P$-case one has the $SU(3)$ relation:
\begin{equation}
m_{P,\tilde{u}\tilde{u}}^2+3m_{P,\tilde{s}\tilde{s}}^2=
4m_{P,\tilde{s}\tilde{u}}^2\,.
\end{equation}
All other relations presented below are derived within the
framework of $SPV\!A$, $U(3)$ QQM.

For the ground $SP$-meson states one has:
\begin{equation}
\label{N}
m_{\sigma,\tilde{u}\tilde{u}}^2-3m_{\pi}^2=
m_{\sigma,\tilde{s}\tilde{u}}^2-3m_{K}^2=m_{\sigma,\tilde{s}\tilde{s}}^2-
3\left(2m_{K}^2-m_{\pi}^2\right)\simeq 4{\cal M}_0^2\,.
\end{equation}
Here and below the dynamic mass ${\cal M}_0\equiv M_0\mid_{m_q=0}$. 
In the $V\!A$-meson sector the relations for ground states look as
follows:
\begin{equation}
m_{a_1,\tilde{u}\tilde{u}}^2-m_{\rho}^2\simeq\frac32\left(
m_{\sigma,\tilde{u}\tilde{u}}^2-m_{\pi}^2\right),\quad
m_{a_1,\tilde{s}\tilde{u}}^2-m_{K^*}^2\simeq\frac32\left(
m_{\sigma,\tilde{s}\tilde{u}}^2-m_{K}^2\right),
\end{equation}
\begin{equation}
m_{a_1,\tilde{s}\tilde{s}}^2-m_{\varphi}^2\simeq\frac32\left
[m_{\sigma,\tilde{s}\tilde{s}}^2-\left(
2m_{K}^2-m_{\pi}^2\right)\right].
\end{equation}
For the excited $V\!A$-meson states one has:
\begin{equation}
m_{a_1',\tilde{u}\tilde{u}}^2-m_{\rho'}^2\simeq\frac32\left(
m_{\sigma',\tilde{u}\tilde{u}}^2-m_{\pi'}^2\right),\quad
m_{a_1',\tilde{s}\tilde{u}}^2-m_{{K^*}'}^2\simeq\frac32\left(
m_{\sigma',\tilde{s}\tilde{u}}^2-m_{K'}^2\right),
\end{equation}
\begin{equation}
m_{a_1',\tilde{s}\tilde{s}}^2-m_{\varphi'}^2\simeq\frac32\left(
m_{\sigma',\tilde{s}\tilde{s}}^2-m_{(\eta)'}^2\right).
\end{equation}



As it was already mentioned, at intermediate energies the correlators
of QQM can be matched~\cite{a5} to
the OPE of QCD correlators~\cite{svz}.
In the large-$N_{c}$ approach
the correlators of color-singlet quark currents are saturated
by narrow meson resonances. In particular, the two-point
correlators are given by the following sums:
\begin{equation}
\Pi^{C}(p^{2})=\int d^{4}x e^{ipx}
\langle T(\bar{q}\Gamma
q(x)\bar{q}\Gamma q(0))
\rangle_{planar}=\sum_{n}\frac{Z_{n}^{C}}{p^{2}+m_{C,n}^{2}}
+D_{0}^{C}+D_{1}^{C}p^{2}, \label{cor1}
\end{equation}
$$
C\equiv S,P,V,A; \qquad
\Gamma=i,\gamma_{5},\gamma_{\mu},\gamma_{\mu}\gamma_{5}; \qquad
D_{0},D_{1}=const.
$$
The last two terms both in the scalar-pseudoscalar and in the
vector-axial-vector channel $D_{0}$ and $D_{1}$ 
are contact terms required for the
regularization of infinite sums. On the other hand
the high-energy asymptotics is
provided~\cite{svz} by the perturbation theory and OPE. Therefrom the
above correlators increase at large $p^{2}$,
\begin{equation}
\Pi^{C}(p^{2})\mid_{p^{2}\rightarrow\infty}\sim
p^{2} \ln\frac{p^{2}}{\mu^{2}}.
\end{equation}
Evidently the
infinite series of resonances with the same quantum numbers
should exist in order to reproduce the perturbative asymptotics.

Meantime the differences of correlators of opposite-parity currents
rapidly decrease at large momenta \cite{ae1, a5} (the chiral limit is
considered below):
\begin{equation}
\left(\Pi^{P}(p^{2})-\Pi^{S}(p^{2})\right)_{p^{2}\rightarrow\infty}\equiv
\frac{\Delta_{SP}}{p^{4}}+{\cal O}\left(\frac{1}{p^{6}}\right),
\qquad \Delta_{SP}\simeq
24\pi\alpha_{s}\langle\bar{q}q\rangle^{2},
\label{SP}
\end{equation}
and~\cite{svz,kr2}
\begin{equation}
\left(\Pi^{V}(p^{2})-\Pi^{A}(p^{2})\right)_{p^{2}\rightarrow\infty}\equiv
\frac{\Delta_{VA}}{p^{6}}-\frac{m_0^2\Delta_{VA}}{p^{8}}+
{\cal O}\left(\frac{1}{p^{10}}\right), \,
\Delta_{VA}\simeq-16\pi\alpha_{s}\langle\bar{q}q\rangle^{2},
\label{VA}
\end{equation}
where $m_0^2=0.8\pm0.2\,\mbox{GeV}^2$~\cite{iz} and we have defined in the
$V,A$ channels
\begin{equation}
\Pi_{\mu\nu}^{V,A}(p^{2})\equiv(-\delta_{\mu\nu}p^{2}+p_{\mu}p_{\nu})
\Pi^{V,A}(p^{2})\,.
\end{equation}
The vacuum dominance hypothesis \cite{svz} in the
large-$N_{c}$ limit is adopted.

Therefore the chiral symmetry is restored at high energies and
the two above differences manifest  genuine order parameters of
CSB in QCD. As they decrease rapidly at large momenta
one can perform the matching of QCD asymptotics
by means of few lowest lying resonances that gives a number of
constraints from Chiral Symmetry Restoration (CSR).

Expanding the meson
correlators~(\ref{cor1}) in
powers of $p^{2}$ one arrives to the CSR Sum Rules. In the
scalar-pseudoscalar case~(\ref{SP}) they read:
\begin{equation}
\sum_{n}Z_{n}^{S}-\sum_{n}Z_{n}^{P}=0\,,\quad
\sum_{n}Z_{n}^{S}m_{S,n}^{2}-\sum_{n}Z_{n}^{P}m_{P,n}^{2}=\Delta_{SP}\,,
\label{pssum}
\end{equation}
and in the vector-axial-vector one~(\ref{VA}) one obtains:
$$
\sum_{n}Z_{n}^{V}-\sum_{n}Z_{n}^{A}=4f_{\pi}^{2}\,,\quad
\sum_{n}Z_{n}^{V}m_{V,n}^{2}-\sum_{n}Z_{n}^{A}m_{A,n}^{2}=0\,,
$$
\begin{equation}
\label{av}
\sum_{n}Z_{n}^{V}m_{V,n}^{4}-\sum_{n}Z_{n}^{A}m_{A,n}^{4}=\Delta_{V\!A}\,,
\quad
\sum_{n}Z_{n}^{V}m_{V,n}^{6}-\sum_{n}Z_{n}^{A}m_{A,n}^{6}=
-m_0^2\Delta_{V\!A}\,.
\end{equation}
The first two relations  are
famous Weinberg Sum Rules, with $f_{\pi}$ being the
pion decay constant. The residues in resonance pole contributions
in the vector and axial-vector correlators have the structure,
\begin{equation}
Z_{n}^{(V,A)}=4f_{(V,A),n}^{2}m_{(V,A),n}^{2}\,,
\end{equation}
with $f_{(V,A),n}$ being defined as corresponding
decay constants.

In the $SP$ case the residues in poles happen to be of different
order of magnitude in logarithms,
\begin{equation}
\frac{Z_{\sigma,\pi}}{Z_{\sigma',\pi'}}={\cal O}\left(
\frac{1}{\ln\!\frac{\Lambda^2}{M_0^2}}\right),
\end{equation}
and when the $\pi-a_1$ mixing is taken into account one derives:
$$
Z_{\pi}\simeq\frac{m_{a_1}^2}{m_{\rho}^2}Z_{\sigma}\simeq
4\frac{\langle\bar qq\rangle^2}{f_{\pi}^2}\simeq
-\frac{N_c\Lambda^4\Delta_{22}(\sigma_1-\sqrt{3}\,\sigma_2)^2}
{12\pi^2m_{\sigma'}^2\sigma_1^2\ln\frac{\Lambda^2}{M_0^2}}\cdot
\frac{m_{a_1}^2}{m_{\rho}^2}\,;
$$
\begin{equation}
Z_{\pi'}=Z_0-Z_{\pi}\simeq Z_{\sigma'}=Z_0-Z_{\sigma}\,,\qquad Z_0\equiv
\frac{N_c\Lambda^4}{2\pi^2}\,.
\end{equation}
The second CSR Sum Rules constraint results in the
estimation for splitting between the $\sigma'$- and
$\pi'$-meson masses,
\begin{equation}
Z_0 (m_{\sigma'}^2 -  m_{\pi'}^2) \simeq 24 \pi\alpha_s
\langle\bar q q\rangle^2. \label{scal}
\end{equation}

In the vector-axial-vector case all residues
are found to be of the same order of magnitude in contrast to the
scalar-pseudoscalar channel. They read:
$$
\widetilde Z_{\pi} 
= 4 f^2_{\pi} \simeq-\frac{N_c\Lambda^4(m_{a_1}^2-m_{\rho}^2)
(6m_{\rho}^2\ln\!\frac{\Lambda^2}{M_0^2}+3\bar\Delta_{11}+
2\sqrt{3}\bar\Delta_{12}+\bar\Delta_{22})}{32\pi^2m_{\rho}^2m_{a_1}^2
m_{a_1'}^2\ln\!\frac{\Lambda^2}{M_0^2}}\,;
$$
$$
Z_{\rho}\simeq\frac{m_{a_1}^2}{m_{a_1}^2-m_{\rho}^2}Z_{\pi}\,,\qquad
Z_{a_1}\simeq\frac{m_{\rho}^2}{m_{a_1}^2-m_{\rho}^2}Z_{\pi}\,;
$$
\begin{equation}
Z_{\rho'}\simeq\frac{Z_1}{m_{\rho'}^2}\,,\qquad
Z_{a_1'}\simeq\frac{Z_1}{m_{a_1'}^2}\,,
\qquad Z_1\equiv\frac{3N_c\Lambda^4}{16\pi^2}\,.
\end{equation}
The relation for $\widetilde Z_{\pi}$ is a
constraint on effective coupling constants of the QQM
$\bar{\Delta}_{kl}$\,.
The first and the second Sum Rules are fulfilled identically.
The third one takes the form:
\begin{equation}
Z_1(m_{a_1'}^2-m_{\rho'}^2)\simeq16\pi\alpha_s\langle\bar{q}q\rangle^2.
\label{3sr}
\end{equation}
The fourth Sum Rule gives in the large-log approach:
\begin{equation}
\label{4sr}
m_{a_1'}^2\simeq m_{\rho'}^2\simeq \frac{m_0^2}{2}\,.
\end{equation}
As it is seen from the Eq.~(\ref{4sr}) the last Sum Rule fails for
QQM with the ground and first excited sets of $V\!A$ mesons.

The relations~(\ref{scal}) and~(\ref{3sr}) constrain the  QQM
parameters following from the OPE. Having as an
input $\Lambda=1000$~MeV and
\begin{equation}
{\cal M}_0=2\sigma_1=320\,\mbox{MeV}\,,\quad
\langle\bar q q\rangle\simeq-\frac{N_c\Lambda^2}{8\pi^2}(\sigma_1-
\sqrt{3}\,\sigma_2)=-(250\,\mbox{MeV})^3,
\end{equation}
one can fix
\begin{equation}
\sigma_1=160\,\mbox{MeV}\,,\qquad \sigma_2=-145\,\mbox{MeV}\,.
\end{equation}
From the QQM relation
\begin{equation}
\label{splits}
m_{a_1'}^2-m_{\rho'}^2\simeq\frac32(m_{\sigma'}^2-m_{\pi'}^2)\simeq
3\sigma_1^2+2\sqrt{3}\,\sigma_1\sigma_2+9\sigma_2^2
\end{equation}
and Eq.~(\ref{scal}),~(\ref{3sr}) one obtains the mass splittings
\begin{equation}
m_{\sigma'}-m_{\pi'}\approx45\,\mbox{MeV}\,,\qquad
m_{a_1'}-m_{\rho'}\approx60\,\mbox{MeV}\,,
\end{equation}
which prove a fast restoration of chiral symmetry. One can
also estimate the required $\alpha_s\approx 0.9$ at one loop. 
It seems to be bearly compartible with perturbative calculations in QCD. 
On the other hand, in the vector channel the
two-loop calculations \cite{iz} 
diminish considerably the value of required strong
coupling constant $\alpha_s\approx 0.5\div 0.6$.

Note that from Eq.~(\ref{splits}) and~(\ref{scal}),~(\ref{3sr}) one may
obtain two independent estimations of the quantity:
\begin{equation}
\frac{m_{a_1'}^2-m_{\rho'}^2}{m_{\sigma'}^2-m_{\pi'}^2}\approx
\left\{
\begin{array}{l}1.5\quad\mbox{from QQM},\\1.8\quad\mbox{from CSR},
\end{array}\right.
\end{equation}
which do not depend on any model parameters. The discrepancy
amounts to 15\%, i.e. it is within the large-$N_c$ approximation.
This shows that saturation of two-point correlators by two
resonances is quite robust.

Finally, in the Appendix  we display the Table with 
our fits for $SPV\!A$ meson masses
and compare them with the corresponding experimental values~\cite{pdg}.
A rather big discrepancy in predicting masses of ground scalar states is
a general problem in phenomenology. In the case of QQM this might
signify that large-$N_c$ corrections are large or the leading-log
approximation does not work well, i.e. the details of confinement
are of importance in this case. Note, however, that the Extended
Chiral Quark Model~\cite{ae1, ae2, ae3} fits better the scalar
sector.

We conclude that the Quasilocal Quark Model reflects
phenomenology of low and intermediate meson physics and passes QCD
tests with the reasonable precision. It must be noticed that in the
conventional NJL model the precision was worse substantially.
Namely, in the SP channel 
one can derive an NJL-model estimation
$$
\frac{Z_{\sigma}m_{\sigma}^2}{24\pi\alpha_s\langle\bar
qq\rangle^2}=\frac{22}{9}\,,
$$
whereas this ratio should be equal 1 from the second CSR Sum Rule
Eq.~(\ref{pssum}).


\bigskip
We express our gratitude to the organizers of the
International Workshop QUARKS 2002 in Novgorod for
hospitality.
This work is supported by Grant RFBR
01-02-17152, INTAS Call 2000 Grant (Project 587), Russian
Ministry of Education Grant E00-33-208, and The Program "Universities
of Russia: Fundamental Investigations" (Grant 992612).

\newpage

\vspace{-3mm}
\begin{table}[tbph]
\caption{The $SPV\!A$, $U(3)$ QQM masses of mesons and their first
excitations (in MeV) for ${\cal M}_0=320$ MeV. $\tilde{u}\equiv u,d$ quarks.}
\vspace{5mm}
\begin{tabular}{|cccccc|c|}
  \hline
  Particle, th. & Particle, exp. & Input & Pred. &
  Experiment & Dif., \% & Case \\
  \hline
  $Singlet$ &  &  &  &  &  &  \\
  $\tilde{u}\tilde{u}$ & $\pi$ & 140 &  & 135-140 &  & $P$ \\
  $\tilde{s}\tilde{u}$ & $K$ & 500 &  & 494-498 &  &  \\
  $\tilde{s}\tilde{s}$ & $\eta$ &  & 570 & 547.30$\pm$0.12 & 4 &  \\
  \hline
  $(Singlet)'$ & $\eta(1295)$ &  & 1300 & 1297.0$\pm$2.8 & $<$1 &  \\
  $\tilde{u}'\tilde{u}'$ & $\pi(1300)$ & 1300 &  & 1300$\pm$100 &  & $P'$ \\
  $\tilde{s}'\tilde{u}'$ & $K(1460)$ & 1400 &  & 1400-1460 &  &  \\
  $\tilde{s}'\tilde{s}'$ & $\eta(1440)$ &  & 1490  & 1400-1470 & 1  &  \\
  \hline
  \hline
  $Singlet$ & $f_0(980)$ &  & 680 & 980$\pm$10 & 31 &  \\
  $\tilde{u}\tilde{u}$ & $a_0(980)$ &  & 680 & 984.8$\pm$1.4 & 31 & $S$ \\
  $\tilde{s}\tilde{u}$ & $K_0^*(960)$ (?) &  & 1080 & 905$\pm$50 (?) & 19 (?) &  \\
  $\tilde{s}\tilde{s}$ & $f_0(1370)$ &  & 1360 & 1200-1500 & (?) &  \\
  \hline
  $(Singlet)'$ & $f_0(1500)$ &  & 1350 & 1500$\pm$10 & 10 &  \\
  $\tilde{u}'\tilde{u}'$ & $a_0(1450)$ &  & 1350 & 1474$\pm$19 & 8 & $S'$ \\
  $\tilde{s}'\tilde{u}'$ & $K_0^*(1430)$ &  & 1440  & 1412$\pm$6 & 2  &  \\
  $\tilde{s}'\tilde{s}'$ & $f_0(1710)$ &  & 1530 & 1715$\pm$7 & 11 &  \\
  \hline
  \hline
  $Singlet$ & $\omega(782)$ &  & 770 & 782.57$\pm$0.12 & 2 &  \\
  $\tilde{u}\tilde{u}$ & $\rho(770)$ & 770 &  & 769.3$\pm$0.08 &  & $V$ \\
  $\tilde{s}\tilde{u}$ & $K^*(892)$ &  & 900 & 892-896 & $<$1 &  \\
  $\tilde{s}\tilde{s}$ & $\varphi(1020)$ & 1020 &  & 1019.417$\pm$0.014 &  &  \\
  \hline
  $(Singlet)'$ & $\omega(1420)$ &  & 1460 & 1419$\pm$31 & 3 &  \\
  $\tilde{u}'\tilde{u}'$ & $\rho(1450)$ & 1460 &  & 1465$\pm$25 &  & $V'$ \\
  $\tilde{s}'\tilde{u}'$ & $K^*(1410)$ &  & 1570 & 1414$\pm$15 & 11 &  \\
  $\tilde{s}'\tilde{s}'$ & $\varphi(1680)$ & 1680 &  & 1680$\pm$20 &  &  \\
  \hline
  \hline
  $Singlet$ & $f_1(1285)$ &  & 1120 & 1281.9$\pm$0.6 & 12 &  \\
  $\tilde{u}\tilde{u}$ & $a_1(1260)$ &  & 1120 & 1230$\pm$40 & 9 & $A$ \\
  $\tilde{s}\tilde{u}$ & $K_1(1400)$ &  & 1470 & 1402$\pm$7 & 5 &  \\
  $\tilde{s}\tilde{s}$ & $f_1(1510)$ &  & 1740 & 1512$\pm$4 & 15 &  \\
  \hline
  $(Singlet)'$ & (?) &  & 1520 & (?) & (?) &  \\
  $\tilde{u}'\tilde{u}'$ & $a_1(1640)$ &  & 1520 & 1640$\pm$40 (?) & 7 & $A'$ \\
  $\tilde{s}'\tilde{u}'$ & $K_1(1650)$ &  & 1630 & 1650$\pm$50 (?) & 1 &  \\
  $\tilde{s}'\tilde{s}'$ & (?) &  & 1730 & (?) & (?) &  \\
  \hline
\end{tabular}
\end{table}

\end{document}